\documentclass[prd,nofootinbib,preprint]{revtex4}
\usepackage{tensor}

\newcommand{\levicivita}{}
\def\levicivita#1#{\tensor#1{\epsilon}}
\usepackage{graphicx}
\usepackage{array}
\usepackage[title]{appendix}
\usepackage{xcolor}
\usepackage{lipsum}
\usepackage{mathtools}
\usepackage{amssymb}
\usepackage{amssymb}
\usepackage{amsmath}
\usepackage{pifont}

\usepackage{epsfig}
\usepackage{makecell}
\usepackage{amsmath}
\usepackage{amsfonts}
\usepackage{amssymb}
\usepackage{mathtools}
\usepackage{url}
\usepackage{hyperref}
\usepackage{subfigure}
\usepackage{hhline}
\usepackage{color}
\usepackage{array,multirow}
\usepackage{bm}
\usepackage{cancel}
\newcommand{\beqa}{\begin{eqnarray}}
\newcommand{\eeqa}{\end{eqnarray}}
\newcommand{\beq}{\begin{equation}}
\newcommand{\eeq}{\end{equation}}

\newcommand{\nn}{\nonumber}
\newcommand{\bmt}{\begin{pmatrix}}
\newcommand{\emt}{\end{pmatrix}}
\usepackage{comment}
\newcommand{\be}{\begin{equation}}
\newcommand{\ee}{\end{equation}}
\newcommand{\bea}{\begin{eqnarray}}
\newcommand{\eea}{\end{eqnarray}}

\begin{document}

\title{Scrutinizing new physics of $B_d\to \phi (\eta^{(')},\pi, \omega)$ decay modes}
\author{Manas K. Mohapatra}
\email{manasmohapatra12@gmail.com}




         
\affiliation{Department of Physics, IIT Hyderabad,
              Kandi - 502285, India }                            

              

\begin{abstract}
We inspect the exclusive hadronic decay modes $ B_d\to \phi (\eta^{(')}, \pi, \omega)$, induced by quark level transition as $b\to d$ $(\Delta S=0)$, in vector like down quark model. As these decay modes insist highly suppressed followed by the predicted branching fraction $\mathcal{O}(10^{-9})$ which reflects to scrutinize physics beyond the standard model. We constrain the new parameter space inferred from experimental limits on leptonic $B_d\to \ell \ell(\ell=e, \mu, \tau)$ and nonleptonic decay modes $B_d \to \eta' \pi^0$ and $B_u\to \rho^- \eta'$. We then check the new physics contributions can have significant impact on the prominent observable so called branching ratio of $B_d\to \phi (\eta^{(')},\pi, \omega)$ processes. 
\end{abstract}
\maketitle

\section{Introduction}
The Standard model (SM) of particle physics, one of the biggest achievements in twentieth century science, encompasses the beauty of fundamental particles and their interactions which is ruled by strong, weak and electromagnetic forces. Despite its spectacular success, it, however,  has some important voids that couldn't filled out such as  matter dominance over antimatter in the present universe, dark matter and dark energy, hierarchy problem, neutrino mass etc. The source of matter-antimatter asymmetry is the violation of combined discrete symmetry of charge conjugation (C) and parity (P) where Cabbibo--Kobayasi--Maskawa (CKM) matrix is the main cornerstone to account for.  Among various indirect searches, the study of B decay modes provide an insight to analyze in the SM, and to explore the possible existence of new physics (NP) beyond it. In one way, it is of interest to study the potential implications in the sector of flavor changing neutral current (FCNC) transition at electroweak scale which mainly occurs at loop level.

In this paper, we will perform the decay modes $ B_d\to \phi (\eta^{(')}, \pi, \omega)$ induced by $b\to d$ quark level transition which are suppressed in the SM indicating an ideal place  to search for the new physics effects. We repredict the CP averaged branching ratios
based on the framework of QCD factorisation approach which includes next to leading order (NLO) contributions where the previous predictions have been done in \cite{Cheng:2009cn, Beneke:2003zv, Beneke:2006hg}.
The upper limit of branching fractions of the aforesaid decay modes are given as \cite{Zyla:2020zbs}
\bea
&&Br(B_d^0\to \phi \eta)< 5 \times 10^{-7}, Br(B_d^0\to \phi \eta')< 5 \times 10^{-7},\nn\\
&&Br(B_d^0\to \phi \pi)< 1.5 \times 10^{-7}, Br(B_d^0\to \phi \omega)< 7 \times 10^{-7}. 
\eea
where as the SM predictions are of $\mathcal{O}(10^{-9})$ with no CPV (CP violation) observables.
Inspired by these discrepancies, we would like to probe in the presence of vector like down quark (VLDQ) model where an extra $SU(2)_L$ singlet down type quark has been added to the SM  and observe the significant impact on the branching fractions. As the SM include 3 generations of quarks but there may be possibility of having a heavier exotic quark in another generation. Such fermion can appear in $E_6$ grand unified theories and models with large extra dimensions. Because of the addition of this particle to the SM particle spectrum, it modifies the CKM matrix, of course not unitary. Thus it implies the FCNC transition at tree level mediated by Z in the down quark sector. We would like to see such effect in our model on the above decay modes.  

Among $b\to d$ quark level transitions are leptonic decays $B_d\to \ell \ell(\ell=e, \mu, \tau)$. The SM and experimental results are shown in the TABLE - \ref{Table:Lep}. The deviations,  between the predictions and experimental values, need modification of the branching fractions in the search of NP scenario.

\begin{table}
\centering
\caption{Branching fractions (leptonic) induced by $b \to d$ transition}
\begin{tabular}{|c|c|c|}
\hline
\hline
Decay processes~&~Predicted Br~&~Experimental values/upper limits \cite{Zyla:2020zbs}\\
\hline
\hline
$B_d^0\to ee$~&~$(2.23 \pm 0.21)\times 10^{-15}$~&~$ <0.83 \times 10^{-8}$  \\
\hline
$B_d^0\to \mu \mu$~&~$(0.95\pm 0.09)\times 10^{-10}$~&~$(1.1\substack{+1.4\\-1.3}) \times 10^{-10}$ \\
\hline
$B_d^0 \to \tau \tau$~&~$(2.00\pm 0.19)\times 10^{-8}$~&~$< 2.1 \times 10^{-3}$\\
\hline
\end{tabular}
\end{table}\label{Table:Lep}

\begin{table}
\centering
\caption{CP av. Branching fractions (non leptonic) induced by $b \to d$ transition}
\label{table:NL}
\begin{tabular}{|c|c|c|c|}
\hline
\hline
Decay modes~&~Our results~&~Previous results \cite{Cheng:2009cn}~&~Experimental value \cite{Zyla:2020zbs}\\
\hline
\hline
$\bar{B}_d \to \pi^0 \eta' $~&~$(0.50 \pm 0.57 \pm 0.03)\times 10^{-6}$~&~$(0.42 \substack{+0.21+0.18 \\-0.09-0.12})\times 10^{-6}$~&~$(1.2 \pm 0.6)\times 10^{-6}$\\
\hline
 $B^-\to \rho^- \eta'$~&~$(6.26 \pm 1.70 \pm 0.34) \times 10^{-6}$~&~$(5.6\substack{+0.9+0.8\\-0.7-0.5})\times 10^{-6}$~&~$(9.7 \pm 2.2) \times 10^{-6}$\\
\hline
\end{tabular}
\end{table}
On the other side, the decay channels having final state meson $\eta'$ play a vital role in the non leptonic $B_d \to \eta ' \pi ^0$ and $B_u \to \rho ^- \eta '$ decay modes where we consider $\eta - \eta'$ mixing effect in the investigation of decay mode having final state meson $\eta$. Now due to the potential deviations between SM and experimental results given in TABLE -\ref{table:NL}, it allows a room to search for  physics beyond the SM. The associated new couplings in the presence of VLDQ model can be constrained by using the experimental limits on both the leptonic as well as non leptonic decay modes. Using the allowed parameter space, we scrutinize the new physics impact on the aforesaid decay processes.

The layout of this work is organized as follows. In section \ref{SMP}, we briefly study the effective Hamiltonian accountable for the quark level transition $b \to d$ for $ B_d\to \phi (\eta^{(')}, \pi, \omega)$ decay modes. We include the discussion of the amplitudes for the abovesaid nonleptonic decay modes along with the numerical results of the so called branching fractions in the SM with all necessary input parameters. In section \ref{NPC}, we then constrain the new parameter space using the existing experimental limits on branching fractions of both leptonic $B_d \to \ell \ell$ and also non leptonic decay modes $B_d \to \eta' \pi^0$ and $B_u\to \rho^- \eta'$, discussed above in the presence of VLDQ model. Next we implement the above model on the decay modes $ B_d\to \phi (\eta^{(')}, \pi, \omega)$ in the new physics scenario by using new coupling paramaters. Finally we bring to an end with brief  summary and conclusion of our results in section \ref{CAD}.

\section{Standard Model Predictions}\label{SMP}

The weak effective Hamiltonian of the decay mode having the
quark level transition $b\to d$, can be written as
\cite{Buchalla:1995vs}
\bea \label{Eff. Hamiltonian}
\mathcal{H} _{eff}=\frac{G_F}{\sqrt{2}} \bigg[V_{ub}V_{ud}^*(C_1(\mu) O_1(\mu)+C_2(\mu)O_2(\mu))-V_{tb}V_{td}^*\sum_{i=3}^{10}C_i(\mu) O_i(\mu)\bigg],
\eea
where the six-dimensional four-quark operators $O_i (i=1,...,10)$ given in the above effective Hamiltonian are specified as below:
\bea
O_1&=(\bar{d}_\alpha \gamma ^\mu Lu_\beta).(\bar{u}_\beta \gamma _\mu L b_\alpha),  &O_6=(\bar{d}_\alpha \gamma ^\mu Lu_\beta).(\sum_q \bar{q}_\beta \gamma _\mu R q_\alpha),\nn\\
O_2&=(\bar{d}_\alpha \gamma ^\mu Lu_\alpha).(\bar{u}_\beta \gamma _\mu L b_\beta), &O_7 =\frac{3}{2}(\bar{d}_\alpha \gamma ^\mu Lu_\alpha).( \sum_q e_q \bar{q}_\beta \gamma _\mu R q_\beta),\nn\\
O_3&=(\bar{d}_\alpha \gamma ^\mu Lu_\alpha).(\sum_q \bar{q}_\beta \gamma _\mu L q_\beta), &O_8=\frac{3}{2}(\bar{d}_\alpha \gamma ^\mu Lu_\beta).( \sum_q e_q\bar{q}_\beta \gamma _\mu R q_\alpha),\nn\\
O_4&=(\bar{d}_\alpha \gamma ^\mu Lu_\beta).(\sum_q \bar{q}_\beta \gamma _\mu L q_\alpha), &O_9=\frac{3}{2}(\bar{d}_\alpha \gamma ^\mu Lu_\alpha).(\sum_q e_q \bar{q}_\beta \gamma _\mu L q_\beta),\nn\\
O_5&=(\bar{d}_\alpha \gamma ^\mu Lu_\alpha).(\sum_q \bar{q}_\beta \gamma _\mu R q_\beta), & O_{10}=\frac{3}{2}(\bar{d}_\alpha \gamma ^\mu Lu_\beta).( \sum_q e_q\bar{q}_\beta \gamma _\mu L q_\alpha),
\eea
where $G_F$ is Fermi coupling constant, all ${V_{ab}}'s(a, b= u, b, s, t)$ are the CKM matrix elements, $e_q$ is the electromagnetic charge of quark field  `q', L (R) is the left (right) handed projection operator, and $\alpha, \beta$ are the color indices. The quark field q runs over active flavors i.e., $q  \epsilon  \{u, d, c, s, b\}$ at the scale $\mu=m_b$. In addition to this, the operators belong to $i=1,2$ are current-current, $i=3,..,6$ are the QCD penguin and $i=7,...,10$ are the EW penguin operators. The corresponding coupling constants so called Wilson coefficients $C_i's (i=1,...,10)$ are used in the next-to-leading order (NLO) at the scale of $O(m_b)$ in order to cancel the $\mu$ dependence of the amplitude.

\section*{\boldmath $B_d \to \phi \eta^{(')}$}
The weak decay amplitude in the presence of QCDF approach \cite{Beneke:2003zv} can be  written in the form as
\bea
\langle \phi \eta ^{(')}|O_{\mathit{i}}|B_d \rangle= \langle \phi \eta ^{(')}|O_{\mathit{i}}|B_d \rangle _{fact}\big[1+\sum r_n \alpha _s^n+\mathit{O}(\frac{\Lambda _{QCD}}{m_b})\big],
\eea
where the factorized matrix element of four-quark operators $\langle \phi \eta ^{(')}|O_{\mathit{i}}|B_d \rangle _{fact}$ includes form factors and decay constants. The second term in the paranthesis involve to higher order contributions which include the QCD effect, more to say, gluon corrections, and the third term directs to power corrections containing troublesome end-point divergence.
The decay mode $B_d \to \phi \eta$ can be produced from the $B_d \to \omega \eta$ process followed by $\omega - \phi$ mixing along with the angle $\delta = 3.3^ \circ$ \citep{Cheng:2009cn,Benayoun:1999fv, Kucukarslan:2006wk, Benayoun:2007cu, Qian:2008px}.
Now the amplitude for the decay mode $B_d \to \omega \eta$ is given by \citep{Cheng:2009cn}
\bea
\mathit{A}(B_d \to \omega \eta)\approx  \frac{1}{2} \lambda _p\bigg[ A_{\omega \eta _q} \big\{ \delta _{pu}(\alpha _2 + \beta _1)+2 \alpha _3^p+ \hat{\alpha}_4^p \big\}+ A_{\eta _q \omega } \big\{\delta _{pu}(\alpha _2 + \beta _1)+2 \alpha _3^p+ \hat{\alpha}_4^p\big\}\bigg],
\eea
where the parameter $\lambda _p$ is the CKM matrix element and is summed over the quark element $p=u, c$.
The required parameters $\alpha_i^p, \beta_i^p$ and$\hat{\alpha}_i^p$ and the factorized matrix elements are given in the Appendix \ref{A}.
The relation between both the branching fractions is given by \citep{Cheng:2009cn}
\bea
Br(B_d \to \phi \eta) = Br(B_d \to \omega \eta) \times \rm \sin ^2 \delta
\eea 
Similarly, we can proceed for the final states $\omega(\phi) \eta'$ where $\eta$ is replaced by $\eta'$ in the previous decay mode.
For our study of B decay modes having final state particle $\eta$ and $\eta'$, we consider $\eta- \eta'$ mixing effect in our study of the physical observable of given decay modes. Due to different flavor states of $\eta^{(')}(\eta_q^{(')}, \eta_s^{(')}$ and $\eta_c^{(')})$, the corresponding decay constants $f_q$ and $f_s$ correlated by a mixing angle $\theta$ which is given by
\bea
\begin{pmatrix}
f_\eta ^q & f _\eta ^s\\
f_{\eta'} ^q & f _{\eta'}^s
\end{pmatrix}
=
\begin{pmatrix}
\cos \theta & - \sin \theta\\
\sin \theta & \cos \theta
\end{pmatrix}
\begin{pmatrix}
f_q & 0\\
0 & f_s
\end{pmatrix},
\eea
where the $\eta_q-\eta_s$ mixing angle $\theta =(39.3 \pm 1.0)^{\circ}$ \cite{Feldmann:1998vh} and the mixing with $\eta_c$ has been neglected 
and the useful parameters $f_q$ and $f_s$ are given by 
\bea
f_q=(1.07\pm 0.02)f_\pi, \hspace*{.5cm} f_s=(1.34\pm 0.06)f_\pi.
\eea

\section*{\boldmath $B_d \to \phi \pi^0$}
Similar to the previous decay mode, the process $B_d \to \phi \pi^-$ also can be produced from the chanel $B_d \to \omega \pi$ in the presence of $\omega - \phi$ mixing effect.
Now the amplitude for the decay mode $B^- \to \pi^- \omega$ is given by\citep{Beneke:2003zv}
\bea
A(B^- \to \pi^- \omega)&& \approx \frac{1}{\sqrt{2}}\lambda _p\bigg[A_{\pi \omega}\big\{\delta _{pu}(\alpha _2 + \beta _2+2 \beta _{S2})+2\alpha _3^p+\alpha _4^p +\frac{1}{2}\alpha _{3,EW}^p-\frac{1}{2}\alpha _{4,EW}^p\nn\\
&&+\beta _3^p+\beta _{3,EW}^p+2 \beta _{S3}^p+2 \beta _{S3,EW}^p \big\} \nn\\
&& +A_{\omega \pi}\big\{ \delta _{pu}(\alpha _1 + \beta _2)+\alpha _4^p +\alpha _{4,EW}^p+\beta _3^p+\beta_{3,EW}^p\big\} \bigg],
\eea
where $\lambda _p$ is the CKM parameter and the other contributions in the above amplitude are given in the Appendix \ref{A}.
Now, the mixing relation between the decay modes $B^- \to \omega ^- \pi$ and $B^- \to \phi \pi^-$ is given by \citep{Cheng:2009cn}
\bea
Br(B^- \to \phi \pi^-)_{{\phi - \omega} \hspace*{0.1cm}mixing}= Br(B^- \to \omega \pi ^-) \times \rm \sin ^2 \delta,
\eea
and the decay amplitude  of $B_d \to \phi \pi^0$ mode is given by the relation as \citep{Beneke:2003zv}
\bea
A(B^-\to \pi^- \phi) = - \sqrt{2} A(B_d \to \phi \pi^0)
\eea
For our calculations of above discussed modes $B_d \to \phi (\pi^0, \eta ^{(')})$, we consider the $\phi-\omega$ mixing angle $\delta= (3.32 \pm 0.09)^ \circ$ from \cite{Ambrosino:2009sc}.
\section*{\boldmath $B_d \to \phi \omega$}
From the effective Hamiltonian (\ref{Eff. Hamiltonian})
the matrix element for the four quark operators is given by
\bea
\big \langle V_1(\lambda _1)V_2(\lambda _2)|(\bar{q}_2 q_3)_{V-A}(\bar{q}_1b)_{V-A}|\bar{B}_d\big \rangle,
\eea
where $\lambda _1, \lambda _2$ are the helicities of the final state vector mesons $V_1$ and $V_2$ respectively.
Now the amplitude of the penguin dominated decay mode $B_d \to \phi \omega$ is expressed as \cite{Beneke:2006hg}
\bea
A(B_d \to \phi \omega)=  A_{\omega \phi} \big[\lambda _p(\alpha _3 ^p -\frac{1}{2}\alpha _{3,EW}^p)\big]
\eea
The details of the parameters used in the above process has been provided in the Appendix \ref{B}.
The helicity amplitudes corresponding to this decay mode are $A_0, A_+$ and $A_-$ and the hierarchy of helicity amplitudes are
\bea
A_0:A_-:A_+=1:\frac{\Lambda _{QCD}}{m_b}:\bigg (\frac{\Lambda _{QCD}}{m_b}\bigg )^2,
\eea
where the transverse amplitudes $A_{+}$ and $A_{-}$ are suppressed relative to the longitudinal one $A_0$.\\
Now in this section, all the discussed non leptonic decay modes include factorized matrix elements $A_{VP}$ and $A_{VV}$ (P = pseudoscalar meson, V = vector meson) as well as the higher order corrections such as vertex corrections, hard spectator interactions, penguin contractions and annihilation contributions. 
Now all the discussed amplitudes can be written in the parameterized form symbolically as 
\bea
A(B \to VP)&&=\lambda _u A_u+ \lambda _c A_c\nn\\
&&=\lambda _c A_c \big[1+ra e^{i(\delta _1-\gamma)}\big],
\eea
where $V=\phi, P= \eta^{(')},\pi$. $\lambda _{u,c}$ are the CKM elements and $A_{u,c}$ are the amplitudes correspond to $u$ and $c$ quark. $a= |\lambda _u/ \lambda _c|$, $r=|A_u/A_c|$, $\gamma$ is the weak phase of  CKM element $V_{ub}$ and the relative strong phase is $\delta _1$.
The formula for CP averaged branching fraction is given by
\bea
\mathfrak{B}=\frac{1}{2} \left[Br(\mathcal{A}_{B_d\to M_1M_2})+Br(\mathcal{\bar{A}}_{B_d\to M_1M_2})\right].
\eea
Now for all the non leptonic decay modes $ B_d\to \phi (\eta^{(')}, \pi)$ can be written as 
\bea
\mathfrak{B}= \frac{p_{cm} \tau _B}{8 \pi m_B^2}|\lambda _cA_c|^2 \big\{1+r^2a^2+2ra \rm \cos \delta _1 \rm \cos \gamma\big\},
\eea
where the center of mass momentum in $B_d$ rest frame is given by 
\bea
p_{cm}=\sqrt{(m_{B_d}^2-(m_{1}+m_{2})^2)(m_{B_d}^2-(m_{1}-m_{2})^2)}\,,
\eea
where $m_1$ and $m_2$ are the masses of final states.
Similarly the CP av. branching fraction for $B_d \to \phi \omega$ is given by 
\bea
\mathfrak{B}_{(B_d \to \phi \omega)}= \frac{\tau _{B_d}p_{cm}}{8 \pi^2 m_B} (|A_0|^2+|A_-|^2+|A_+|^2).
\eea
Here CP av. branching fraction correspond to all the indivisual helicity amplitude can be written similar to the expressions in $B_d \to VP$ decay mode.

Now the CP averaged branching ratio can be calculated for the considered nonleptonic $ B_d\to \phi (\eta^{(')}, \pi, \omega)$ decay modes. For the numerical predictions of the CP averaged branching ratio, we use the input parameters given in S4 scenario of QCDF approach \cite{Beneke:2003zv}. 
The Wilson coefficients in NDR scheme at NLO are taken from \cite{deGroot:2003ms} at $m_b$ scale and the relevant input parameters are given in the Table \ref{Input parameters}.
Many studies have been done in the Ref. \cite{Beneke:2003zv,Cheng:2009cn,Beneke:2006hg,Zhang:2008bz,Cheng:2014rfa}. We repredict SM values of the CP averaged branching fractions of $ B_d\to \phi (\eta^{(')}, \pi, \omega)$ decay modes which are given in the TABLE \ref{table:SMNL} along with the previous results.
Here the first theoretical error correspond to the uncertainties occured due to quak masses, form factor, decay constants, Gegenbauer moments, the wave function of $B_d^0$ meson and  $\phi -\omega$ mixing angle where as the parameters due to weak annhilation and hard spectator interactions are lumped into the second uncertainty. As per the ref. \citep{Cheng:2009cn} we assign $0.1$ and $20^ \circ$ uncertainties to the annihilation parameters $\rho _A$ and $\phi _A$ respectively.
\begin{table}
\centering
\caption{SM predictions of CP av. branching fractions (non leptonic) induced by $b \to d$ transition}
\label{table:SMNL}
\begin{tabular}{|c|c|c|c|}
\hline
\hline
Decay modes~&~Our results ~&~Previous results \cite{Beneke:2003zv,Bao:2008hd}~&~Expt. values \cite{Zyla:2020zbs}\\
\hline
\hline
$B_d \to \phi \eta$~&~$(1.18 \pm 0.84\pm 0.03)\times 10^{-9}$~&~$0.001\times 10^{-6}$~&~$< 5 \times 10^{-7}$\\
\hline
 $B_d\to \phi \eta'$~&~$(2.26 \pm 1.8 \pm 0.09) \times 10^{-9}$~&~$0.003\times 10^{-6}$~&~$<5 \times 10^{-7}$\\
\hline
$B_d\to \phi \pi^0$~&~$(6.91 \pm 1.23 \pm 0.03) \times 10^{-9}$~&~$0.004\times 10^{-6}$~&~$<1.5 \times 10^{-7}$\\
\hline
$B_d\to \phi \omega$~&~$(3.16 \pm 1.23 \pm 0.006) \times 10^{-9}$~&~$0.0017\times 10^{-6}$~&~$<7 \times 10^{-7}$\\
\hline
\end{tabular}
\end{table}

\begin{table}
\centering
\caption{Input parameters used in the numerical analysis} \label{Input parameters}
\begin{tabular}{ |p{14cm}| p{1cm}| p{1cm}| } 
\hline
\hline
Running quark masses and coupling constants: & Ref. \\
\hline
$G_F = 1.166 \times 10^{-5}$ GeV$^{-2}$; \hspace{0.1cm} $\alpha _{em}=1/129$  & \citep{Zyla:2020zbs}  \\
$\alpha _s(M_Z)=0.1185$; \hspace{0.1cm} $\tau _{B_d}=(1.52 \pm 0.004) \times 10^{-12}\hspace{0.1cm} s$ & \citep{Zyla:2020zbs}  \\
$m_b(m_b)=4.2 \hspace{0.1cm} \rm GeV$; \hspace{0.1cm} $m_c(m_b)=0.91\hspace{0.1cm} \rm GeV$; \hspace{0.1cm} $m_c^{\rm pole}/m_b^{\rm pole}=0.3$ & \citep{Cheng:2009cn} \\
$m_u(2 \hspace{.01cm} \rm GeV)=2.15 \pm 0.15$ \hspace{0.01cm} MeV; \hspace{0.2cm}$m_d(2 \hspace{.01cm}\rm GeV)=4.7 \pm 0.2$ \hspace{0.01cm} MeV & \cite{Lu:2018cfc} \\
$m_s(2\hspace{.01cm} \rm GeV)=93.8 \pm 1.3\pm 1.9$ \hspace{.01cm} MeV & \cite{Lu:2018cfc} \\
\hline
\hline
CKM parameters: & \\
\hline
$V_{ub}=(3.82\pm 0.24)\times 10^{-3}$; $V_{ud}=0.97370 \pm 0.00014$; $V_{cb}=(41 \pm 1.4)\times 10^{-3}$ & \cite{Zyla:2020zbs}\\ 
$V_{cd}=0.221\pm 0.004$; $V_{td}=(8.0\pm 0.3)\times 10^{-3}$; $V_{tb}=1.013\pm 0.030$ & \cite{Zyla:2020zbs} \\
$\gamma = (72.1 \substack{+4.1\\-4.5})^\circ$; $\sin 2\beta _d=0.699 \pm 0.017$ & \cite{Zyla:2020zbs} \\
\hline
\hline
Form factors and decay constants: & \\
\hline
$F_{B\to \eta}(0)=0.168 \substack{+0.041 \\-0.047}$; \hspace{0.1cm} $F_{B\to \eta'}(0)=0.130 \substack{+0.036\\ -0.032}$; & \cite{Duplancic:2015zna} \\
 $F_{B\to \pi}(0)=0.21 \pm 0.07$; $A_{B\to \rho}(0)=0.356 \pm 0.042$;   & \cite{Gubernari:2018wyi,Straub:2015ica} \\
 $A^0_{B\to \omega}(0)=0.328\pm 0.048$;\hspace*{.1cm}$A^1_{B\to \omega}(0)=0.243\pm 0.031$;\hspace*{.1cm}$V_{B\to \omega}(0)=0.304\pm 0.038$; $f_\omega=(187\pm 5)$ MeV; $f^ \perp _\omega=(151\pm 9)$ MeV  & \cite{Straub:2015ica,Cheng:2009cn} \\
$f_\eta ^q=107$ MeV; $f_\eta ^s=-112$ MeV; $f_{\eta'}^q=89$ MeV;  $f_{\eta'}^s=137$ MeV& \cite{,Cheng:2009cn} \\
$f_{B_d}=(190.5\pm 1.3)$ MeV; $f_\pi=(130.2 \pm 1.4)$ MeV; &\cite{Aoki:2019cca,Gubernari:2018wyi}\\
$f_\rho=(216 \pm 3)$ MeV; $f_\rho^ \perp=(165 \pm 9)$ MeV & \cite{Cheng:2009cn}\\
 $f_\phi=(215 \pm 5)$ MeV; $f_\phi^ \perp=(186 \pm 9)$ MeV & \cite{Cheng:2009cn}\\
\hline
\hline
Gegenbauer moments: & \\
\hline
$a_1^\phi=0$; \hspace{0.1cm} $a_2^\phi=0.18\pm0.08$; \hspace{0.1cm} $a_1^{\perp , \phi}=0$; \hspace{0.1cm} $a_2^{\perp , \phi}=0.14\pm0.06$ & \cite{Cheng:2009cn} \\
$a_1^\rho=0$; \hspace{0.1cm} $a_2^\rho=0.15\pm0.07$; \hspace{0.3cm}$a_1^{\perp , \rho}=0$; 
\hspace{0.1cm} $a_2^{\perp , \rho}=0.14\pm0.07$ & \citep{Cheng:2009cn}\\
$a_1^\pi=0$; \hspace{0.1cm} $a_2^\pi=0.25\pm 0.15$ & \citep{Cheng:2009cn} \\
$a_1^\omega=0$; \hspace{0.1cm} $a_2^\omega=0.15\pm 0.07$; $a_1^{\perp, \omega}=0$; $a_2^{\perp, \omega}=0.14\pm 0.06$; $\lambda _B=300 \pm 100$ MeV& \citep{Cheng:2009cn}\\
\hline
\hline
Annihilation and hard spectator parameters: & \\
\hline
PP mode: $\rho _A = 1.1$; $\phi=-50^\circ$ & \citep{Cheng:2009cn} \\
PV mode: $\rho _A = 0.87$; $\phi=-30^\circ$& \citep{Cheng:2009cn} \\
VP mode: $\rho _A = 1.07$; $\phi=-70^\circ$;$X_H=2.4 \pm 0.024$ & \citep{Cheng:2009cn} \\

\hline
\end{tabular}
\end{table}

\newpage
 \section{New-Physics Contributions} \label{NPC}
In the standard model, flavor changing neutral current (FCNC) occurs at loop level and provide a  very strong suppression because of the intermediate light quark contributions. Therefore it would be more challenging to explore the NP beyond the SM. In this work we include a self-consistent framework where a minimal extension of the standard model in other words enlarging the matter sector having an extra iso singlet vector-like down quark represent to this where Z boson is mediated with FCNC transition at tree level. Now due to the addition of down type quark, the interaction lagrangian for Z boson in the weak eigen state basis can be represented as \cite{Deshpande:2004xc}
\bea
L_Z= -\frac{g}{2 c_W} \big[\bar{U}_L^0\gamma ^ \mu U_L^0- \bar{D}_L^0 \gamma ^ \mu D_L^ 0 -2s_W^2(Q_u \bar{U}^0\gamma ^ \mu U^0 + Q_d \bar{D}^0 \gamma ^ \mu D^0 +Q_d \bar{D}^{0'}\gamma ^\mu D^{0'}\big]Z_ \mu,
\eea
where $Q_{u,d}$ are the electric charges of up and down type quarks. The up type quark $U^0$ and the down type quark $D^0$ are embeded in the SM three generations of quarks and the additional down type quark is given by $D^{0'}=d^{0'}$. Now because of the extension of down type quark, the down quark matrix and the up quark matrix can be diagonalized by $4 \times 4$ and $3 \times 3$ matrix respectively.
So the corresponding interaction Lagrangian mediated by Z, is given as \cite{Giri:2003jj}
$$\mathcal{L}_Z=\frac{g}{2 c _W}\big[\bar{\mathit{U}}_{L\mathit{i}} \gamma ^ \mu \mathit{U}_{L\mathit{i}} -\bar{D} _{L \alpha} U_{\alpha \beta} \gamma ^\mu D_{L \alpha} -2 s^2_W  J_{em}^\mu\big] Z_ \mu,$$
where $i$ ($\alpha, \beta$) denote the generation indices for up (down)-type and $L$ indicate for the left chiral particles. Here the focus point is the second term where the matrix $U_{\alpha \beta}$ is a $4 \times 4$ matrix and the expression is represented as
\bea\label{Unitarity}
U_{\alpha \beta}=\sum_{\mathit{i}=\mathit{u, c, t}}V_{\alpha \mathit{i}}^\dagger V_{\mathit{i}\beta}=\delta_{\alpha \beta}-V_{4 \alpha}^*V_{4 \beta}.
\eea
And this is the distinctive feature of this model. The corresponding CKM matrix for the charge current interaction would be $V= V_u^{L\dagger}V_d^L$ which is a $3 \times 4$ pseudo matrix. It is usually different from the CKM matrix present in the standard model. Since $U_{\alpha \beta}(\alpha, \beta = b, d, s)\neq 0$, it motivates to study FCNC mediated by Z boson at tree level\cite{Buras:1998vd, Alok:2014yua, Mohapatra:2019wcm, Giri:2004wn}. 

Now the non unitrary matrix V arrises due to the addition of the extra 4th quark to the SM sector.
Thus it provides a new signal to scrutinize the physics beyond the SM. Now we constrain the  new parameter space arising due to both leptonic as well non leptonic modes.
\subsection{ \textbf{Constraint from leptonic $B_d \to \ell^+ \ell^-(\ell =e, \mu, \tau)$, and non leptonic modes $B_d \to \eta' \pi^0$ and $B_u\to \rho^- \eta'$}:}
The leptonic modes $B_d\to \ell^+\ell^- (\ell = e, \mu, \tau)$ are suppressed in the SM, still it can be investigated  in the new physics scenario in the presence of VLDQ model.
The branching fraction of $B_d\to \ell^+\ell^-$ in $Z$ mediated VLDQ model is given by \cite{Chen:2010aq}
\bea\label{Br VLDQ}
\mathfrak{B}_{(B_d\to \ell^+\ell^-)}=\frac{G_F^2\alpha ^2m_{B_d}m_\ell ^2f_{B_d}^2 \tau_{B_d}}{16\pi ^3} |V_{tb}V_{td}^*|^2\sqrt{1-4(\frac{m_\ell ^2}{m_{B_d}^2})}\left|C_{10}^{\rm tot}\right |^2,
\eea
where 
\bea
C_{10}^{\rm tot}=C_{10}-\frac{\pi}{\alpha}\frac{U_{bd}}{V_{tb}V_{td}^*}\,.
\eea
Here the term $U_{bd}$ is the coupling parameter when b quark talks to d quark in the presence of mediating Z particle and because of quark mixing it may behave as a complex with weak phase $\phi _d$.
Now the amplitude of the non leptonic $B^- \to \rho^- \eta'$ process is given by\cite{Beneke:2003zv},
\bea
-\sqrt{2}\mathcal{A}_{B^-\to \rho \eta '}&=&A_{\rho\eta '_q}\big[\delta _{pu}(\alpha _2-\beta_2+2\beta_{S2})+2\alpha_3^p+\alpha_4^p+\frac{1}{2}\alpha_{3,EW}^p-\frac{1}{2}\alpha_{4,EW}^p\nn\\
&+&\beta_3^p+\beta_{3,EW}^p+2\beta_{S3}^p+2\beta_{S3,EW}^p\big]\nn\\
&+&\sqrt{2}A_{\rho \eta'_s}\big[\delta_{pu}\beta_{S2}+\alpha_3^p-\frac{1}{2}\alpha_{3,EW}^p+\beta_{S3}^p+\beta_{S3,EW}^p\big]\nn\\
&+&\sqrt{2}A_{\rho \eta_c}\big[\delta_{pc}\alpha_2+\alpha_3^p\big]\nn\\
&+&A_{\eta'_q\rho}\big[\delta_{pu}(\alpha_+\beta_2)+\alpha_4^p+\alpha_{4,EW}^p+\beta_3^p+\beta_{3,EW}^p\big].
\eea
And the amplitude of the decay mode $B_d\to \pi^0 \eta'$ is given as \cite{Beneke:2003zv},
\bea
-2\mathcal{A}_{\bar{B}_d \to \pi ^0 \eta'}&=&A_{\pi\eta_q}\big[\delta _{pu}(\alpha _2-\beta_1-2\beta_{S1})+2\alpha_3^p+\alpha_4^p+\frac{1}{2}\alpha_{3,EW}^p-\frac{1}{2}\alpha_{4,EW}^p\nn\\
&+&\beta_3^p-\frac{1}{2}\beta_{3,EW}^p-\frac{3}{2}\beta_{4,EW}^p+2\beta_{S3}^p-\beta_{S3,EW}^p-3\beta_{S4,EW}^p\big]\nn\\
&+&\sqrt{2}A_{\pi \eta_s}\big[-\delta _{pu}\beta_{S1}+\alpha_3^p-\frac{1}{2}\alpha_{3,EW}^p+\beta_{S3}^p-\frac{1}{2}\beta_{S3,EW}^p-\frac{3}{2}\beta_{S3,EW}^p\big]\nn\\
&+&\sqrt{2}A_{\pi \eta_c}\big[\delta_{pc}\alpha_2+\alpha_3^p\big]\nn\\
&+&A_{\eta_q\pi} \big[\delta_{pu}(-\alpha_2-\beta_1)+\alpha_4^p-\frac{3}{2}\alpha_{3,EW}^p-\frac{1}{2}\beta_{4,EW}^p+\beta_3^p-\frac{1}{2}\beta_{3,EW}^p-\frac{3}{2}\beta_{4,EW}^p\big],
\eea
where the above decay mode amplitudes are multiplied by the CKM element $\lambda _p $ and summed over $p=u, c$.  The required parameters are given in the Appendix \ref{A}. 
Now the effective Hamiltonian corresponding to new interaction describing quark lvel transition $b\to d$ can be represented as
$$\mathcal{H}_{eff}^Z=-\frac{G_F}{\sqrt{2}}V_{tb}V_{td}^*\big[\tilde{C}_3O_3+\tilde{C}_7O_7+\tilde{C}_9O_9\big],$$
where the new Wilson coefficients at the $M_Z$ scale are given as \cite{Atwood:2003tg, Deshpande:2003nx}
\bea \label{new couplings}
\tilde{C}_3(M_Z)&=&\frac{1}{6}\frac{U_{bd}}{V_{tb} V_{td}^*},\nn\\
\tilde{C}_7(M_Z)&=&\frac{2}{3}\frac{U_{bd}}{V_{tb} V_{td}^*}\sin^2\theta _W\,, \nn\\
\tilde{C}_9(M_Z)&=&-\frac{2}{3}\frac{U_{bd}}{V_{tb} V_{td}^*}(1-\sin^2\theta _W)\,
\eea
and the new Wilson coefficients at $m_b$ scale can be found in \cite{Mawlong:2008mb}.
From the unitary condition (\ref{Unitarity}), we get
\bea
\lambda _u +\lambda _c + \lambda _t =U_{bd}\,.
\eea
Now the amplitude in the presence of new physics can be parameterized as,

\bea \label{NP-amplitude-parametrize}
\mathcal{A}&=& \lambda_u \mathcal{A}_u +\lambda_c \mathcal{A}_c-U_{bd} \mathcal{A}_{NP} \nn\\
&=& \lambda_c A_c\Big[1+a r e^{\mathit{i}(\delta_1-\gamma)}-a'r'e ^{\mathit{i}(\delta' + \phi _d)}\Big]\,,
\eea
where 
\bea
a=|\frac{\lambda _u}{\lambda _c}|, \hspace{3mm} r=|\frac{A_u}{A_c}|, \hspace{3mm}
a'=|\frac{U_{bd}}{\lambda _c}|,  \hspace{3mm} r' =|\frac{A_{NP}}{A_c}|.
\eea
Here, $\gamma$, the weak phase arises from the CKM matrix element $V_{ub}$. $\delta_1$  and $\delta'$ are the relative strong phase of $A_u$ and $A_{NP}$ respectively with $A_c$ where the subscript u and c quark correspond the amplitude involved to up and charm quark. Here the  new coupling parameter $U_{bd}$ may have complex phase $\phi_d$. From the amplitude given in Eq.(\ref{NP-amplitude-parametrize}), the CP averaged branching ratio can be written as
\bea\label{NP BR}
\mathfrak{B} &=&\frac{\tau _{B_d} p_c}{8 \pi m_{B_d}^2} |\xi _c A_c|^2\bigg[\mathcal{G}+2r a\cos \delta_1 \cos \gamma -2 r' a'\cos \delta' \cos \phi _d  \nn \\&& -2 rr' aa' \cos(\delta_1 -\delta') \cos (\gamma + \phi _d)\bigg] \,,     
\eea
where $\mathcal{G}= 1+(ra)^2+(r'a')^2$.
Now combining both the leptonic and non leptonic modes with the given experimental and theoretical values from TABLE - \ref{Table:Lep} and \ref{table:NL},
the new parameter space $U_{bd}- \phi_d$ within $1 \sigma$ limit is represented in FIG.\ref{Fig:123}\,. Now the new parameter ranges are shown below.
\bea \label{consrnt}
&1.60\times 10^{-6}\leq |U_{bd}|\leq 2.22\times 10^{-4}, \hspace{0.5cm} 92.70^ {\circ}\leq \phi _d \leq 360^ {\circ},\nn\\
&1.32 \times 10^{-5} \leq |U_{bd}|\leq 1.69\times 10^{-4}, \hspace{0.5cm} 211.55^ {\circ}\leq \phi _d \leq 360^ {\circ}.
\eea

\begin{figure}
\caption{The allowed region of new coupling parameter space $U_{bd} - \phi_d$ arised from the branching fractions of both  leptonic $B_d\to \ell ^+ \ell ^-(\ell =e, \mu, \tau)$, and non leptonic $B_d \to \eta' \pi^0$ and $B_u\to \rho^- \eta'$  processes.}
\centering
\includegraphics[scale=0.4]{{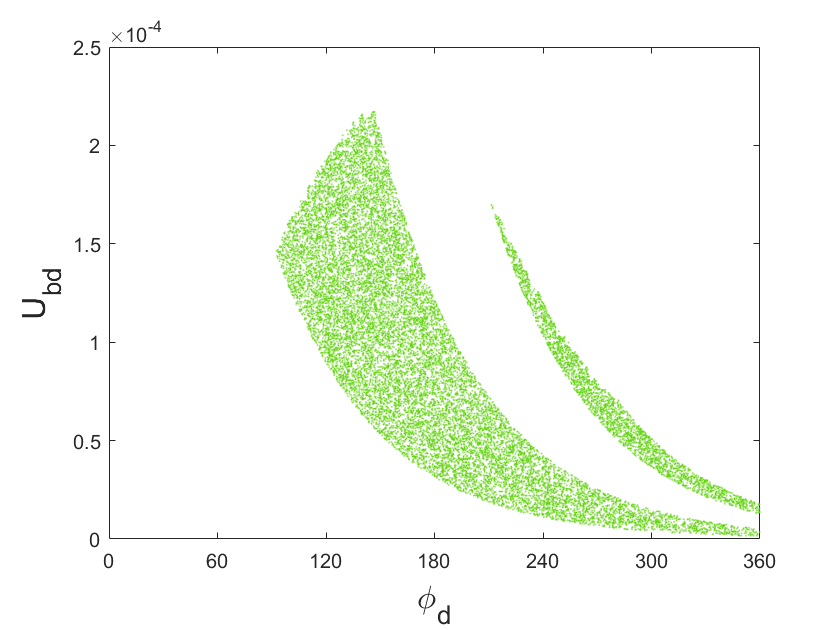}} 
\label{Fig:123}
\end{figure}

\subsection{Impact on the non leptonic modes:}
\subsubsection*{$B_d \to \phi \eta$:}
\begin{figure}[htb] 
\caption{$B_d \to \phi \eta$: Variation of CP averaged branching ratio (in the units of $10^{-8}$) of with ($\mathit{i}$) some benchmark points of $U_{bd}$  as $2\times 10^{-5}$ (Blue), $4\times 10^{-5}$ (Purple) and $6\times 10^{-5}$ (Green) with the new weak phase $\phi_d$  where the dashed black line represents to the SM value with the red dot-dashed line along with the yellow region denote its $1\sigma$ uncertainty (left panel), and with ($\mathit{ii}$) all possible values of $U_{bd}$ and $\phi_d$ (right panel).}
\centering
\includegraphics[scale=0.87]{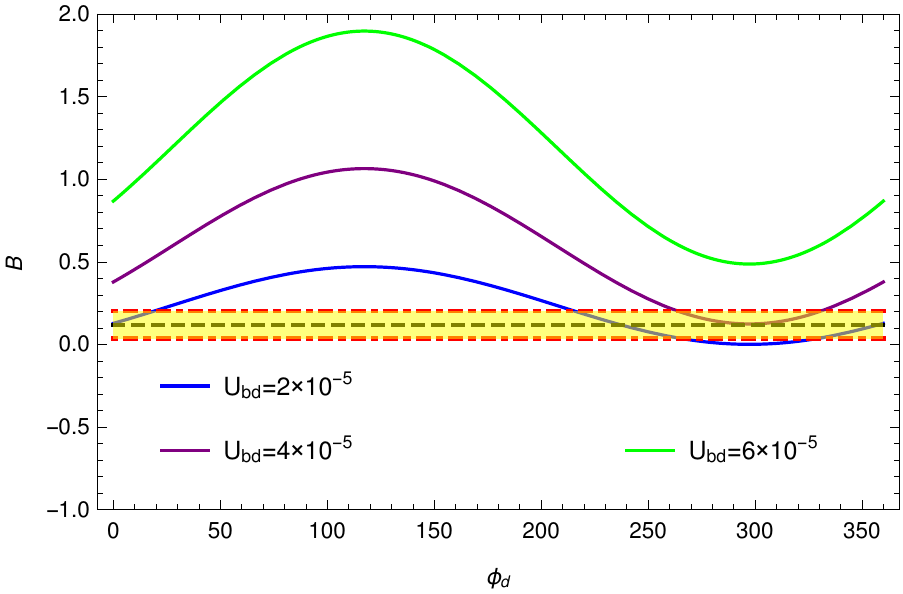}
\quad
 \includegraphics[scale=0.60]{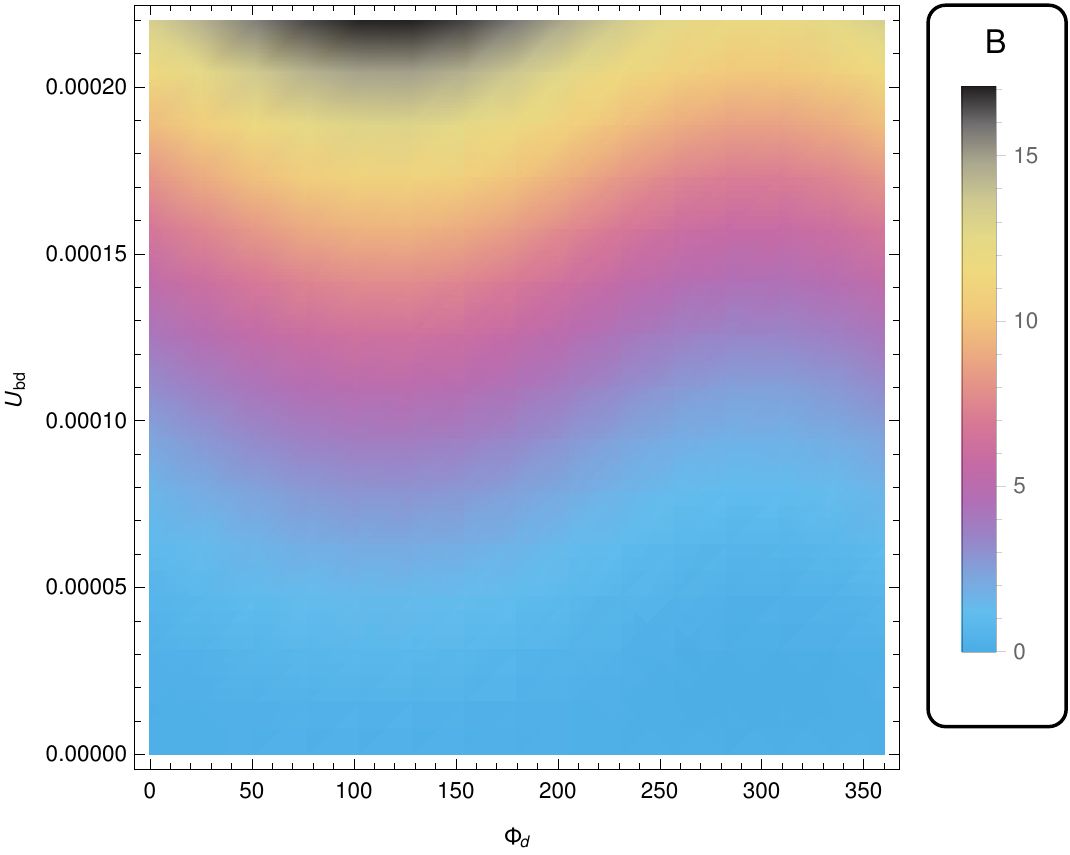} 
\label{NL-1}
\end{figure}

Using the allowed parameter space, we present the variation of CP averaged branching ratio $\mathcal{B}$ with the weak phase $\phi _d$ by considering three benchmark entries of the parameter $|U_{bd}|$ as $2\times 10^{-5}, 4 \times 10^{-5}$ and $6\times 10^{-5}$ given in FIG.-\ref{NL-1} in the left panel. The black dotted central line correspond to the SM value where as the red dot-dashed lines shaded with the yellow color represent its $1\sigma$ uncertainty. From this figure we see that during the variation of the weak phase for the benchmark value $|U_{bd}|=2 \times 10^{-5}$ (blue line), the observable has significantly deviated from its SM contribution in the region $0 \leq \phi_d \leq 240^ \circ$.
Similarly one can also observe that for the other two benchmark entries (purple and green line), the  CP av. branching fraction has also effective contribution from its standard model prediction. Additionally, in the presence of the constraint parameters $U_{bd}$ and $\phi_d$ from the two regions (\ref{consrnt}), the observable has constructive contribution to the standard model value. In the right panel we have shown the variation of the observable (in the units of $10^{-8}$) in the presence of all possible entries of the parameter $|U_{bd}|$ and with the weak phase $\phi _d$. 
\subsubsection*{$B_d \to \phi \eta'$:}
\begin{figure}[htb]
\centering
\includegraphics[scale=0.87]{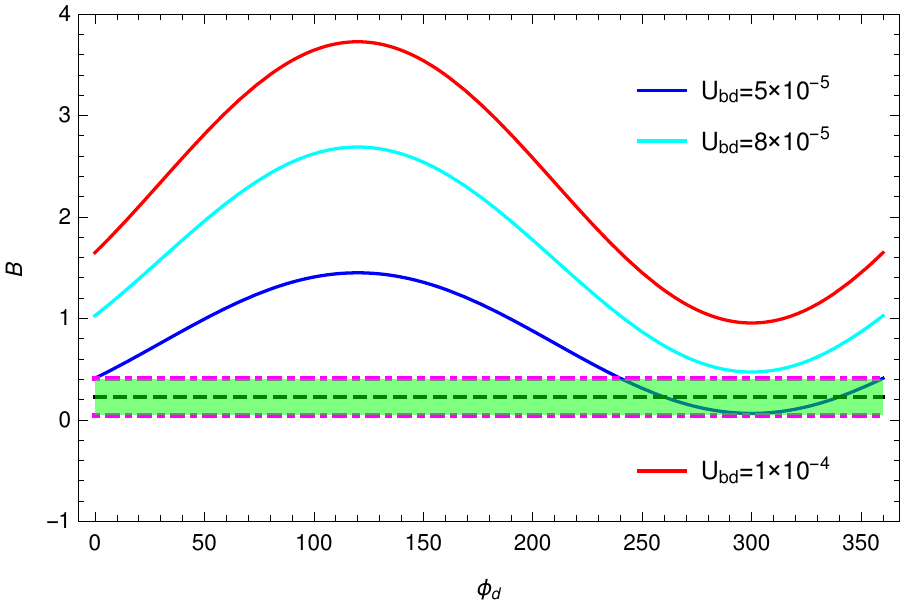}
\quad
 \includegraphics[scale=0.60]{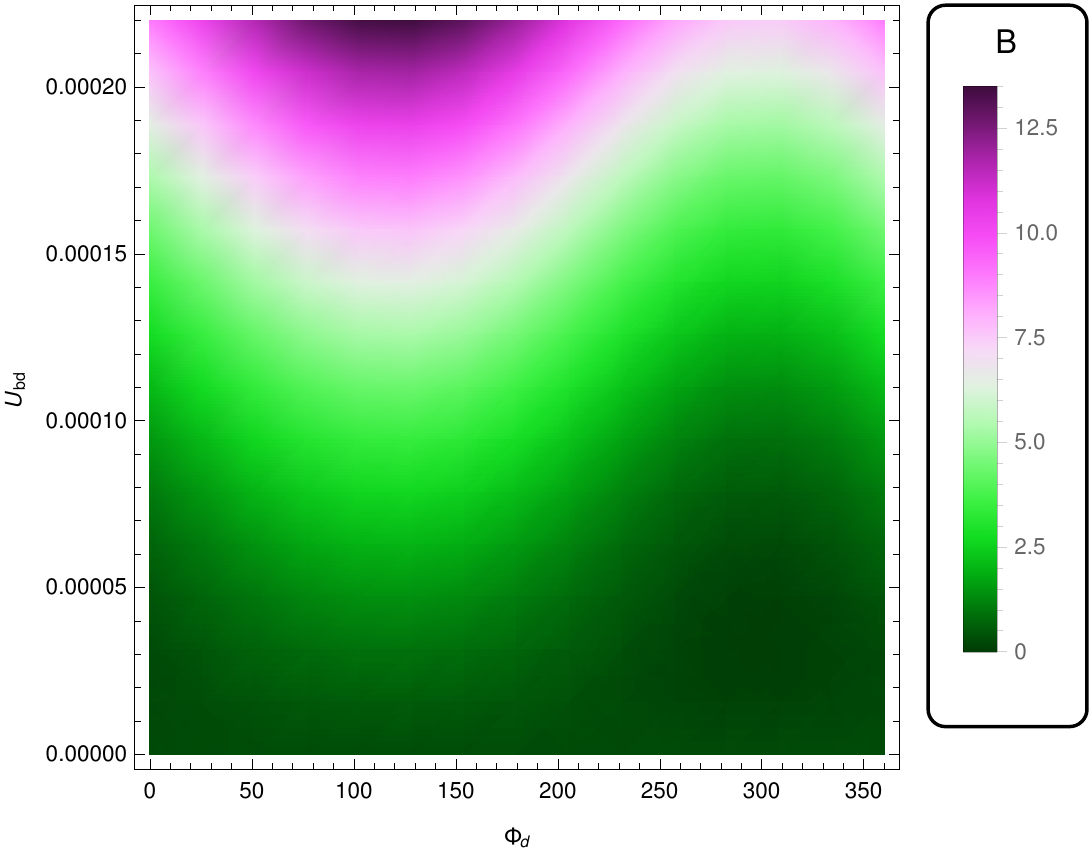} 
\caption{$B_d \to \phi \eta'$: Variation of CP averaged branching ratio (in the units of $10^{-8}$)  with the new weak phase $\phi_d$ ($\mathit{i}$) in the presence of benchmark points of $U_{bd}$ as $5\times 10^{-5}$ (Blue), $8\times 10^{-5}$ (Cyan) and $1\times 10^{-4}$ (Red) (left panel) where the dashed black line represent to the SM value along with $1 - \sigma$ error  (green region), and with ($\mathit{ii}$) all possible values of $U_{bd}$ and $\phi_d$ (right panel).}
\label{NL-2}
\end{figure}

Similarly in FIG.- \ref{NL-2}, we display the impact of the new coupling parameter on variation of the CP averaged branching ratio with the weak phase $\phi _d$ for the decay mode $B_d \to \phi \eta'$. We study with three different values of the parameter $|U_{bd}|$ whose entries are $5\times 10^{-5}, 8\times 10^{-5}$ and $1\times 10^{-4}$, and these entries correspond to the line with blue, cyan and red color in the given figure (left panel) respectively. Also the black dotted line and the magenta dot-dashed line (shaded with green color)  denote the SM value and its $1\sigma$ error respectively. Now from this one can notice clearly in the left panel that the observable in the presence of the benchmark point $|U_{bd}|$ correspond to blue line has deviated towards the $1\sigma$ range of SM line in the region $ 240^ \circ \leq \phi _d \leq 360 ^ \circ$  while it has contributions above to the SM in the other region. Moreover it could be significantly enhanced from the SM value in the presence of the other two benchmark points while in the constraint regions of the NP parameters given in eq. (\ref{consrnt}), it has remarkable contributions. Additionally the right panel shows the impact of all the entries of the new physics parameter on the CP av. branching ratio (in the units of $10^{-8}$).

\subsubsection*{$B_d \to \phi \pi^0$:}
\begin{figure}[htb]
\caption{$B_d \to \phi \pi^0$: Variation of CP averaged branching ratio (in the units of $10^{-8}$)  with ($\mathit{i}$) three benchmark points of $U_{bd}$ as $5\times 10^{-5}$ (Magenta), $9\times 10^{-5}$ (Green) and $2\times 10^{-4}$ (Cyan) with the new weak phase $\phi_d$ (left panel) where the dashed black line represent to the SM value along with shaded region of $1 - \sigma$ uncertainty, and with ($\mathit{ii}$) all possible values of $U_{bd}$ and $\phi_d$ (right panel).}
\centering
\includegraphics[scale=0.87]{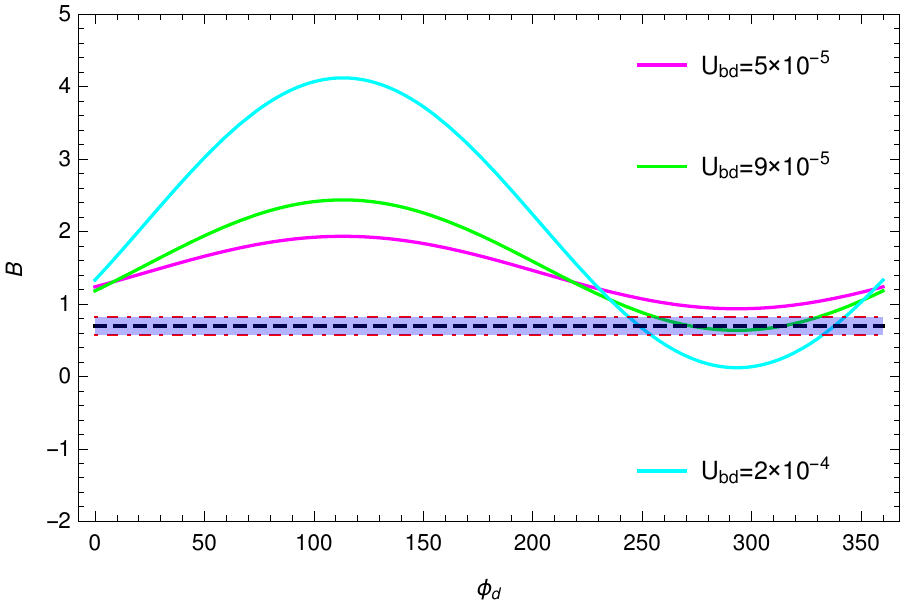}
\quad
 \includegraphics[scale=0.60]{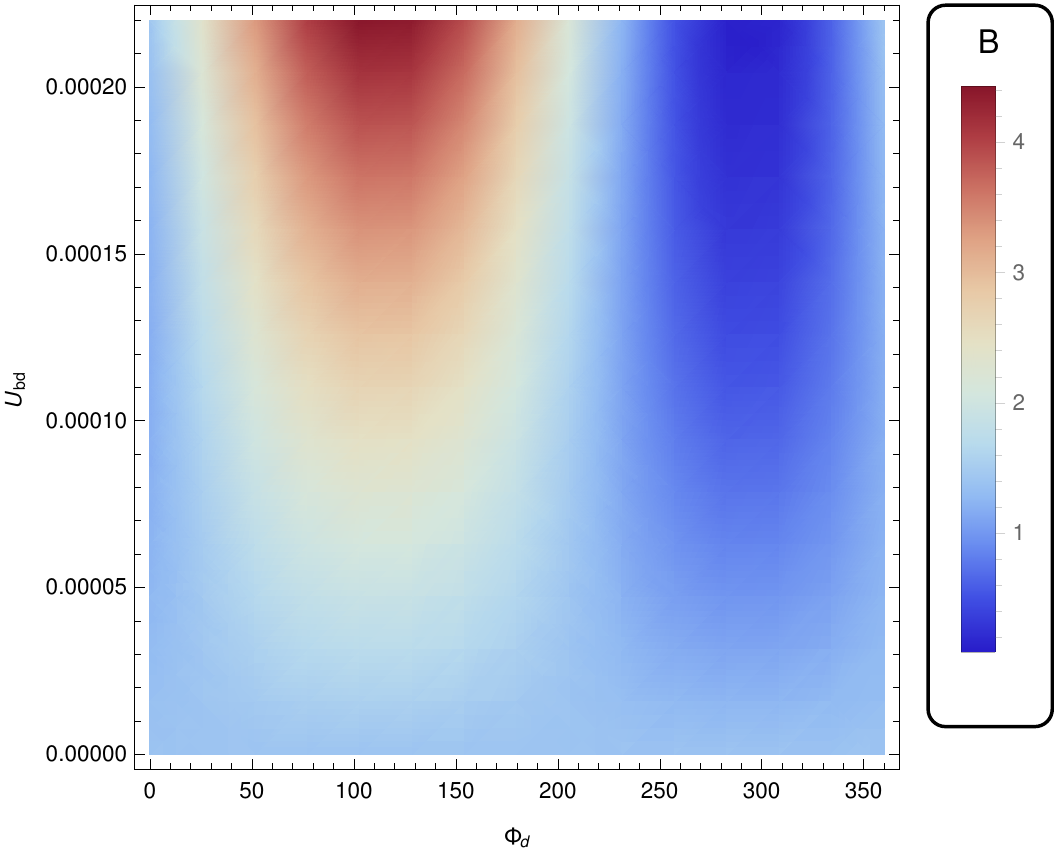} 
\label{NL-31}
\end{figure}

Here we investigate the CP av. branching ratio of the process $B_d \to \phi \pi^0$ with respect to the weak phase $\phi _d$. The Z - b - d coupled parameter $U_{bd}$ has important contributions to the observable in the presence of NP scenario  and is displayed in the left panel of FIG. - \ref{NL-31} with three benchmark inputs. The black dotted line corresponds to the SM prediction where as the light blue colored region along with red dot-dashed line denote its $1\sigma$ deviation. With the 3 different inputs of $|U_{bd}|$, we observe that the observable has significantly deviated from the SM result. Moreover we get more deviations while increasing the coupling parameter $|U_{bd}|$.
For the ranges of $0^\circ\leq \phi_d\leq 230^\circ$, the CP av. branching ratio could be effectively deviated from the stadard model value. In addition to this, the right panel having the variation of the new parameter with all entries along with the phase $\phi _d$, the observable displays all its deviations. However the observable has significant impact in the regions of the new coupling parameters given in eq. (\ref{consrnt}).

\subsubsection*{$B_d \to \omega \phi$:}
\begin{figure}[htb]
\centering
\includegraphics[scale=0.87]{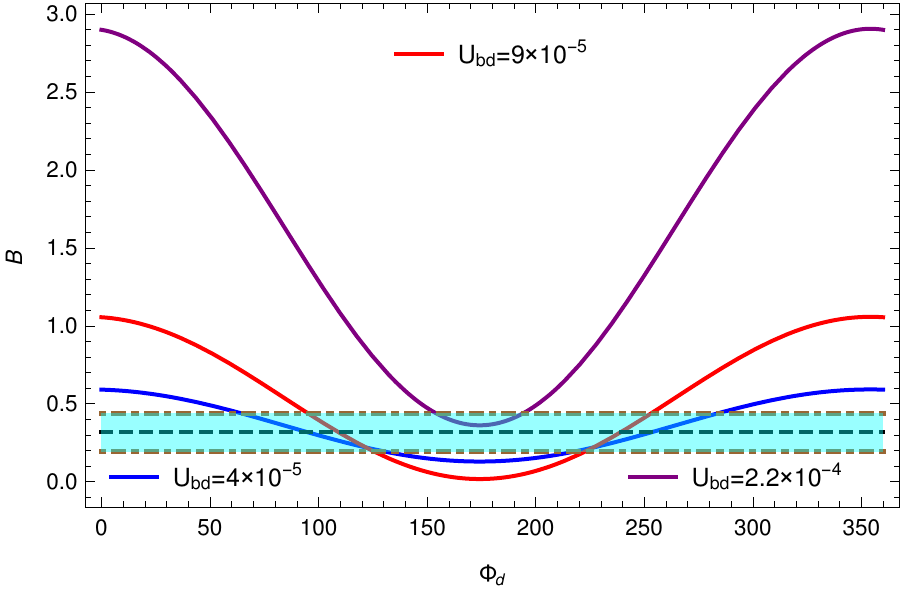}
\quad
 \includegraphics[scale=0.60]{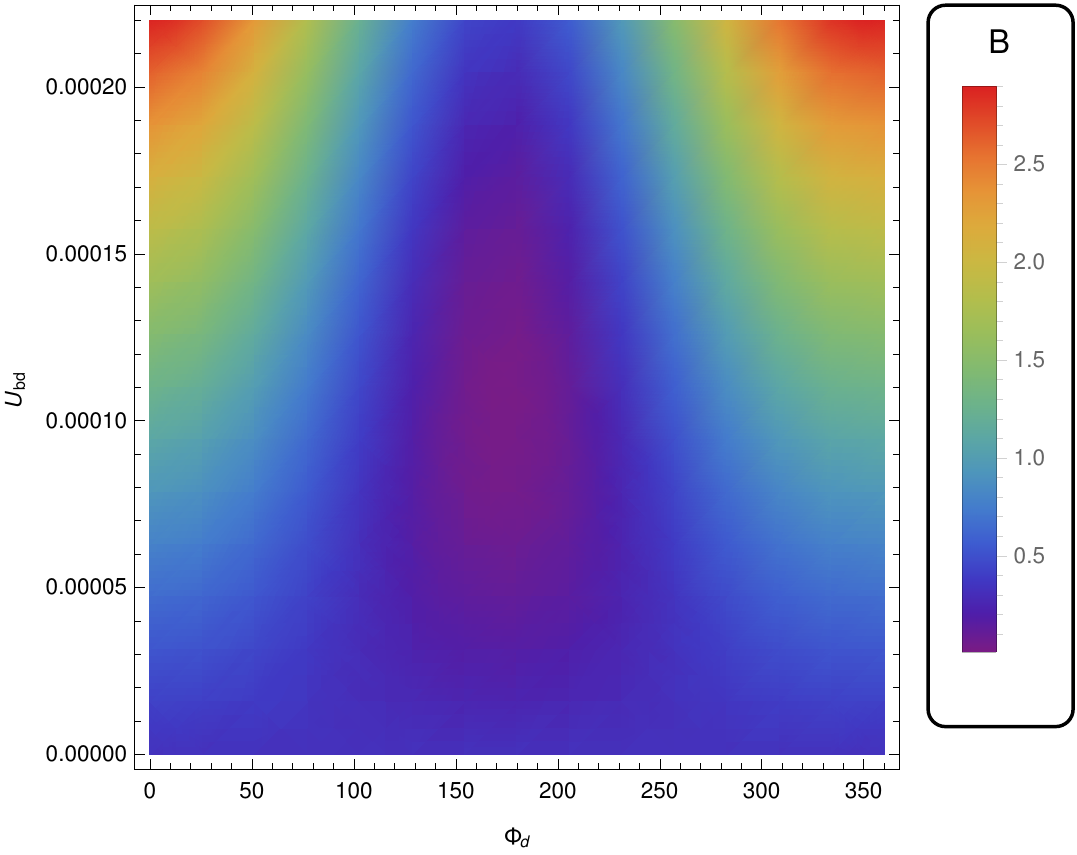} 
\caption{$B_d \to \phi \omega$: Variation of CP averaged branching ratio (in the units of $10^{-8}$)  with ($\mathit{i}$) some benchmark points of $U_{bd}$ as $4\times 10^{-5}$ (Blue), $9\times 10^{-5}$ (Red) and $2.2\times 10^{-4}$ (Purple) with the new weak phase $\phi_d$ (left panel) where the dashed black line represent to the SM value, with ($\mathit{ii}$) all possible values of $U_{bd}$ and $\phi_d$ (right panel).}
\label{NL-4}
\end{figure}

Now in the study of $B_d \to \omega \phi$  process, the new physics parameter has contributed effectively to the variation of CP av. branching ratio with respect to the phase $\phi_d$. The corresponding FIG. \ref{NL-4} (left panel) represents that the new physics in the presence of  the benchmark values of $|U_{bd}|$, the above decay mode has effective contributions from its standard model value. The region of central black dotted line with dot-dashed grey line shaded with cyan color provides $1\sigma$ uncertainty to the SM. Taking a careful observation to the contribution corresponding to the input value of $|U_{bd}|=4 \times 10^{-5}$  $(9 \times 10^{-5})$, the observable in the range of $0 ^ \circ \leq  \phi _d\leq 95^ \circ, 255^ \circ \leq \phi _d\leq 360^ \circ$ ($0 ^ \circ \leq  \phi _d\leq 110^ \circ, 240^ \circ \leq \phi _d\leq 360^ \circ$) has less deviations while in the span of $0^ \circ \leq \phi _d \leq 155 ^ \circ$ and  $200 ^ \circ \leq \phi _d \leq 360 ^ \circ$ from its $1\sigma$ range, it has deviated more effectively than the other two contributions. Similar to other decay modes discussed above, we vary the observable with all the contributions of new physics parameters in the right panel of the given figure. Furthermore, in the regions of the sizeable parameters given in eq. (\ref{consrnt}), the physical observable has significant impact in the presence of the VLDQ model.

\section{Conclusion}\label{CAD}
We have scrutinized the decay modes of $B_d\to \phi (\eta^{(')},\pi, \omega)$, induced by $b\to d$ quark level  transition beyond the standard model.
In the new physics scenario, we have considered vector-like down quark  model where a new quark generation has been added to the SM and consequently it provides the interaction of $Z$ mediated FCNC  at the tree level. 
As the leptonic modes $B_d \to \ell \ell (\ell = e, \mu, \tau)$ and the non leptonic modes $B_d\to \pi \eta'$ and $B_u \to \rho^- \eta'$ have descrepancies between the SM and experimental values, we investigated in the presence of VLDQ model. In the presence of new physics, we constrained the region of parameter space associated with  the interactions``$Z-b-d$'' at tree level.  We found that with the sizeable new coupling parameter $U_{bd}$, considered from both leptonic as well as nonleptonic modes, it provides significant contributions to $B_d\to \phi (\eta^{(')},\pi, \omega)$ processes.
\acknowledgments 

MKM would like to thank to Department of Science and Technology(DST)- Inspire Fellowship division, Government of India for financial support through ID No - IF160303.  MKM  would like to acknowledge Prof. Anjan Giri for his support and
useful discussions.

\bibliography{Bd-phi-eta}
\appendix

\begin{appendices}
\section{\underline {The parameters used in the nonleptonic $B \to PP, PV, VP$ decay modes} \label{A}} \label{A}
We need the factorized matrix elements for the decay mode $B\to M_1M_2$ which are given by
\bea
A_{M_1M_2}= i\frac{G_F}{\sqrt{2}}
                 \begin{cases}
                 m_B^2F_0^{B\to M_1}(0)f_{M_2}; &   {\rm for}  \hspace*{.2cm}  M_1=M_2=\rm Pseudoscalar,\\
                 -2m_V\epsilon ^*_{M_1}.p_B A_0^{B\to M_1}(0)f_{M_2};  &   {\rm for}  \hspace*{.2cm} M_1=\rm Vector, M_2=\rm Pseudoscalar,\\
                 -2m_V\epsilon ^*_{M_1}.p_B F_+^{B\to M_1}(0)f_{M_2};  &   {\rm for}  \hspace*{.2cm} M_1=\rm Pseudoscalar, M_2=\rm Vector.
                 \end{cases}
\eea 
The form factors $F_+$ and $F_0$ denote pseudoscalar mesons, $A_0$ stands for vector meson where as $f_P$ and $F_V$ denote the decay constant for pseudoscalar and vector meson respectively.\\
The expressions of flavor operators in QCD factorisation process are given as follows:
\bea
 \alpha _1 (M_1M_2)&=& a_1 (M_1M_2),\nn\\
 \alpha _2 (M_1M_2)&=& a_2 (M_1M_2),\nn\\
 \alpha _3^p (M_1M_2)&=& 
                 \begin{cases}
                 a_3 (M_1M_2) - a_5 (M_1M_2); &   {\rm for}  \hspace*{.2cm}  M_1M_2=PP,VP,\\
                 a_3 (M_1M_2) + a_5 (M_1M_2); &    {\rm for}  \hspace*{.2cm} M_1M_2=VV,PV,
                 \end{cases}\nn\\
\alpha _4^p (M_1M_2)&=&
                 \begin{cases}
                 a_4^p (M_1M_2) + r_\chi ^{M_2} a_6^p (M_1M_2); &   {\rm for}  \hspace*{.2cm}    
                 M_1M_2=PP,PV,\\
                 a_4^p (M_1M_2) - r_\chi ^{M_2} a_6^p (M_1M_2); &    {\rm for}  \hspace*{.2cm} 
                 M_1M_2=VV,VP,
                 \end{cases} \nn\\
\alpha _{3,EW}^p (M_1M_2)&=& 
                 \begin{cases}
                 a_9 (M_1M_2) - a_7 (M_1M_2); &   {\rm for}  \hspace*{.2cm}  M_1M_2=PP,VP,\\
                 a_9 (M_1M_2) + a_7 (M_1M_2); &    {\rm for}  \hspace*{.2cm} M_1M_2=VV,PV,
                 \end{cases} \nn\\
\alpha _{4,EW}^p (M_1M_2)&=&
                 \begin{cases}
                 a_{10}^p (M_1M_2) + r_\chi ^{M_2} a_8^p (M_1M_2); &  {\rm for}  \hspace*{.2cm}    
                 M_1M_2=PP,PV,\\
                 a_{10}^p (M_1M_2) - r_\chi ^{M_2} a_8^p (M_1M_2); &  {\rm for}  \hspace*{.2cm} 
                 M_1M_2=VV,VP,
                 \end{cases}
\eea 
where 
\bea
a_i^p(M_1M_2) = \bigg(C_i + \frac{C_{i \pm 1}}{N_c}\bigg)N_i(M_2) + \frac{C_{i\pm1}}{N_c}\frac{C_F \alpha _s}{4 \pi} \bigg[V_i (M_2) + \frac{4 \pi ^2}{N_c} H_i (M_1M_2)\bigg] + P_i^p (M_2),
\eea
and $\hat{\alpha} _{4}^p= \alpha _{4}^p + \beta _3^p$ with the superscript $p=u,c$. The details of the parameter $\beta_3^p$ is given below. The quantity $N_i (M_2)$ also reads as
\bea
   N_i (M_2) = \begin{cases}
                 0; &   i=6,8,\\
                 1; &   {\rm otherwise},
                 \end{cases}
\eea
and $i$ runs from 1 to 10. The lower and upper sign correspond to even and odd values of $i$,  where as $C_i's$ and $C_F$ are the Wilson coefficients and the color factor (with $N_c$ = 3) respectively. The relevant contributions $V_i(M_2)$ and $H_i(M_1M_2)$ are vertex corrections and hard spectator interactions where as the term $P_i ^p(M_1 M_2)$ shows as penguin contractions. The explicit expressions are given below.
\begin{itemize}
\item \underline{Vertex corrections}: 
\end{itemize}
\bea
       V_i(M_2)= 
                 \begin{cases}
                 \int _{0}^{1}dx {\rm \Phi}_{M_2}(x) \big[12  {\rm ln} \frac{m_b}{\mu}-18 +g(x)\big]; &   {\rm for} \hspace*{.2cm} i=1-4, 9, 10, \nn\\
                 \int _{0}^{1}dx {\rm \Phi}_{M_2}(x) \big[-12  {\rm ln} \frac{m_b}{\mu}+6 -g(1-x)\big]; &   {\rm for} \hspace*{.2cm} i=5,7, \nn\\
                 \int _{0}^{1}dx {\rm \Phi}_{m_2}(x) \big[-6 +h(x)\big]; &   {\rm for} \hspace*{.2cm} i=6,8,
                 \end{cases}
\eea
where 
\bea
g(x) &=&3\bigg(\frac{1-2x}{1-x} {\rm ln}x - i \pi \bigg)\nn\\
    &+&\bigg[2 Li_2(x) - {\rm ln}^2 x + \frac{2 {\rm ln} x}{1-x} -(3 + 2 \pi i) {\rm ln}x - (x \leftrightarrow 1-x)\bigg],\nn\\
    h(x)&=&2 Li_2 (x) -{\rm ln}^2x- (1+2 \pi i) {\rm ln}x -(x \leftrightarrow 1-x).
\eea
The terms so called $\Phi_{P,V}(x)$ and $\Phi _{p,v}(x)$ given in the above expessions are leading twist and twist-3 light-cone distribution amplitudes, respectively \cite{Beneke:2003zv}. 
\begin{itemize}
\item \underline {Hard spectator interactions}:
\end{itemize}
\bea
H_i(M_1M_2)=\frac{B_{M_1M_2}}{A_{M_1M_2}}\frac{m_B}{\lambda _B}\int _0^1 dx \int _0^1 dy \bigg[\frac{\Phi _{M_2}(x)\Phi _{M_1}(y)}{\bar{x}\bar{y}}+ r_ \chi ^{M_1} \frac{\Phi _{M_2}(x)\Phi _{m_1}(y)}{x\bar{y}}\bigg]
\eea
for $i=1-4,9,10,$
\bea
H_i(M_1M_2)=-\frac{B_{M_1M_2}}{A_{M_1M_2}}\frac{m_B}{\lambda _B}\int _0^1 dx \int _0^1 dy \bigg[\frac{\Phi _{M_2}(x)\Phi _{M_1}(y)}{x\bar{y}}+ r_ \chi ^{M_1} \frac{\Phi _{M_2}(x)\Phi _{m_1}(y)}{\bar{x}\bar{y}}\bigg]
\eea
for $i=5,7$ and $H_i(M_1M_2)=0$ for $i=6,8$
where we consider $\lambda _B$= 300 MeV.
\begin{itemize}
\item \underline{Penguin contractions}:
\end{itemize}
These terms at the order of $\alpha_s$ are given as
\bea
P_4^p(M_2)&=&\frac{C_F \alpha_s}{4 \pi N_c} \bigg\{C_1\bigg[\frac{4}{3} {\rm ln} \frac{m_b}{\mu}+\frac{2}{3}-G_{M_2}(s_p)\bigg]+C_3\bigg[\frac{8}{3}{\rm ln}\frac{m_b}{\mu} +\frac{4}{3}-G_{M_2}(0)-G_{M_2}(1)\bigg]\bigg\}\nn\\
&+&(C_4+C_6)\bigg[\frac{4n_f}{3}{\rm ln} \frac{m_b}{\mu} -(n_f-2)G_{M_2}(0)-G_{M_2}(s_c)-G_{M_2}(1)\bigg]\nn\\
&-&2C_{8g^{\rm eff}}\int _0^1\frac{dx}{1-x}\Phi_{M_2}(x),
\eea
\bea
P_6^p(M_2=P)&=&\frac{C_F\alpha_s}{4\pi N_c}\bigg\{C_1 \bigg[\frac{4}{3}{\rm ln} \frac{m_b}{\mu}+\frac{2}{3}-\hat{G}_{M_2}(s_p)\bigg]+C_3\bigg[\frac{8}{3}{\rm ln} \frac{m_b}{\mu}+\frac{4}{3}-\hat{G}_{M_2}(0)-\hat{G}_{M_2}(1)\bigg] \nn\\
&+&(C_4+C_6)\bigg[\frac{4n_f}{3}{\rm ln} \frac{m_b}{\mu}-(n_f-2)\hat{G}_{M_2}(0)-\hat{G}_{M_2}(s_c)-\hat{G}_{M_2}(1)\bigg]-2C_{8g}^{\rm eff} \bigg\},\nn\\
P_6^p(M_2=V)&=&-\frac{C_F\alpha_s}{4\pi N_c}\bigg\{C_1 \bigg[\hat{G}_{M_2}(s_p)\bigg]+C_3\bigg[\hat{G}_{M_2}(0)-\hat{G}_{M_2}(1)\bigg]\nn\\
&+&(C_4+C_6)\bigg[(n_f-2)\hat{G}_{M_2}(0)+\hat{G}_{M_2}(s_c)+\hat{G}_{M_2}(1)\bigg]\bigg\},\nn\\
P_8^p(M_2=P)&=&\frac{\alpha}{9 \pi N_c}\bigg\{(C_1+N_cC_2)\bigg[\frac{4}{3}{\rm ln}\frac{m_b}{\mu} +\frac{2}{3}-\hat{G}_{M_2}(s_p)\bigg]-3C_{7\gamma}^{\rm eff}\bigg\},\nn\\
P_8^p(M_2=V)&=&-\frac{\alpha}{9 \pi N_c}(C_1+N_cC_2)\hat{G}_{M_2}(s_p),
\eea
\bea
P_{10}^p&=&\frac{\alpha}{9 \pi N_c}\bigg\{(C_1+N_cC_2)\bigg[\frac{4}{3}{\rm ln}\frac{m_b}{\mu} +\frac{2}{3}-G_{M_2}(s_p)\bigg]-3C_{7\gamma}^{\rm eff} \int _0^1 \frac{dx}{1-x}\Phi _{M_2}(x)\bigg\},
\eea
where $n_f = 5$, $s_u=(\frac{m_u}{m_b})^2 \approx 0$ and $s_c=(\frac{m_c}{m_b})^2$. The parameters so called $\alpha _s$ and $\alpha$ are strongand EM coupling constants respectively. The functions $G_{M_2}(s)$ and $\hat{G}_{M_2}(s)$ are defined in \cite{Beneke:2003zv}. In addition to this, the power suppressed weak annihilation contributions are given by
\begin{itemize}
\item \underline{Annihilation contribution}:
\end{itemize}
\bea
\beta_i ^p (M_1M_2)=\frac{i f_B f_{M_1}f_{M_2}}{A_{M_1M_2}} b_i^p,
\eea
where
\bea
&& b_1=\frac{C_F}{N_c^2} C_1 A_1^i, \hspace*{.2cm} b_3= \frac{C_F}{N_c^2} \big[C_3A_1^i + C_5 (A_3 ^i + A_3^f) + N_c C_6 A_3 ^3\big],\nn\\
&& b_2 =\frac{C_F}{N_c^2} C_2 A_1^i, \hspace*{.2cm} b_4= \frac{C_F}{N_c^2} \big[C_4 A_1^i + C_6 A_2 ^f\big],\nn\\
&& b_{3,EW}^p=\frac{C_F}{N_c^2}\big[C_9 A_1^i + C_7(A_3^i + A_3^f) + N_c C_8 A_3^i\big],\nn\\
&& b_{4,EW}^p=\frac{C_F}{N_c^2}\big[ C_{10}A_1^i + C_8(A_2^i \big].
\eea
Here the expressions of A are given as:\\
\underline{Case - I ($M_1=M_2=P$)}:
\bea
&&A_1^i\approx A_2^i\approx 2 \pi \alpha_s \big[9(X_A-4+\frac{\pi^2}{3})+r_\chi ^{M_1}r_\chi ^{M_2} X_A^2\big],\nn\\
&&A_3^i\approx 6 \pi \alpha_s (r_\chi^{M_1}-r_\chi^{M^2})\big(X_A^2-2X^A+\frac{\pi^2}{3}\big),\nn\\
&&A_3^f \approx 6\pi \alpha_s (r_\chi^{M_1}+r_\chi^{M^2})(2X_A^2-X_A),\nn\\
&&A_1^f =A_2^f=0.
\eea
\underline{Case - II ($M_1=V, M_2=P$)}:
\bea
&&A_1^i\approx -A_2^i\approx 6 \pi \alpha_s \big[3(X_A-4+\frac{\pi^2}{3})+r_\chi ^{M_1}r_\chi ^{M_2} (X_A^2-2X_A)\big],\nn\\
&&A_3^i\approx  6 \pi \alpha_s \bigg[-3r_\chi^{M_1}\big(X_A^2-2X^A+\frac{\pi^2}{3}+4\big)+r_\chi ^{M_2}\big(X_A^2-2X_A+\frac{\pi^2}{3}\big)\bigg],\nn\\
&&A_3^f \approx 6\pi \alpha_s\big[3r_\chi^{M_1}(2X_A-1)(2-X_A)-r_\chi ^{M_2}(2X_A^1-X_A)\big],\nn\\
&&A_1^f =A_2^f=0.
\eea
\underline{Case - III ($M_1=P, M_2=V$)}:
\bea
&&A_1^i\approx -A_2^i\approx 6 \pi \alpha_s \big[3(X_A-4+\frac{\pi^2}{3})+r_\chi ^{M_2}r_\chi ^{M_1} (X_A^2-2X_A)\big],\nn\\
&&A_3^i\approx  6 \pi \alpha_s \bigg[-3r_\chi^{M_2}\big(X_A^2-2X^A+\frac{\pi^2}{3}+4\big)+r_\chi ^{M_1}\big(X_A^2-2X_A+\frac{\pi^2}{3}\big)\bigg],\nn\\
&&A_3^f \approx -6\pi \alpha_s\big[3r_\chi^{M_2}(2X_A-1)(2-X_A)-r_\chi ^{M_1}(2X_A^1-X_A)\big],\nn\\
&&A_1^f =A_2^f=0.
\eea
$A_n^{i, f}:$ n = 1, 2 and 3 correspond to the the operator structure $(V-A)(V-A), (V-A)(V+A)$ and $(S-P)(S+P)$ respectively where as the superscripts - ($i,f$) denote for the gluon emission from initial and final states. The chiral factor $r_\chi$ is given by 
\bea
r_\chi ^P(\mu) = \frac{2m_P^2}{m_b(\mu)(m_1+m_2)(\mu)}, \hspace*{.2cm} r_\chi ^V(\mu) = \frac{2m_V}{m_b(\mu)} \frac{f_V^{\perp}(\mu)}{f_V}.
\eea
The end point divergence that has been used, can be given as \bea
X_A= {\rm ln} \frac{m_B}{\Lambda _{QCD}} (1+ {\rm \rho_A} \exp ^{i \phi _A}),
\eea
where ${\rm \rho_A}$ and $\phi_A$ can be found from \cite{Cheng:2009cn}.
\begin{center}
\begin{tabular}{ |c|c|c|c| } 
\hline
Modes & $\rm \rho _A$ & $\rm \phi _A$ \\
\hline
$B\to PP$ & 1.10 & $-50^{\circ}$ \\ 
\hline
 $B\to PV$ & 0.87 & $-30^{\circ}$ \\ 
\hline
$B\to VP$ & 1.07 & $-70^{\circ}$ \\ 
\hline
\end{tabular}
\end{center}

\section{\underline {The parameters used in the nonleptonic $B \to VV$ decay mode} \label{B}}
For the decay process $B \to VV$ the helicity amplitudes depend upon the factorized matrix elements as \cite{Beneke:2006hg}
\bea
X_{B_d \to V_1,V_2}= \big \langle V_2|(\bar{q}_2q_3)_{V-A}|0 \big \rangle \big \langle V_1|(\bar{q}_1b)_{V-A}|\bar{B}_d \big \rangle,
\eea
where the form factor and the decay constants are defined by 
\bea
\big \langle V(p, \epsilon ^*)|\bar{q}\gamma _ \mu q'|0\ big \rangle =&& - i f_Vm_V\epsilon _ \mu ^*, \nn\\
\big \langle V(p, \epsilon ^*)\bar{q}\gamma _ \mu (1-\gamma _5)b|\bar{B}_d (p_B)\big \rangle =&& -\epsilon _\mu ^* (m_B+m_V)A_1 ^{B_d V}(q^2)+ (p_B+p)_\mu (\epsilon ^* .p_B)\frac{A_2 ^{B_dV}(q^2)}{m_B +m_V}\nn\\
&& + q_\mu (\epsilon ^* p_B)\frac{2m_V}{q^2}[A_3^{B_dV}(q^2)-A_0^{B_dV}(q^2)]\nn\\
&& -i \mathcal{E} _{\mu \nu \alpha \beta}\epsilon ^{* \nu} p_B^\alpha p^ \beta
\frac{2V^{B_dV}(q^2)}{m_B +m_V},
\eea
where $q=p_B -p$. The expressions of the helicity amplitudes are given as
\bea
X^0_{V_1V_2}=i \frac{G_F}{\sqrt{2}}m_B^2f_{V_2}A_0^{B\to V_1}(0), \hspace*{.3cm} X_{V_1V_2}^{\pm}= \frac{G_F}{\sqrt{2}}m_Bm_2f_{V_2}F_{\pm}^{B \to V_1}(0),
\eea
where the form factor $F_\pm ^{B \to V_1}$ is defined as
\bea
F_{\pm}^{B \to V_1}(q^2)=(1+\frac{m_1}{m_B})A_1^{B \to V_1}(q^2) \mp (1-\frac{m_1}{m_B})V^{B\to V_1}(q^2).
\eea
The assembled form of the coefficients $a_i$ are given as
\bea
a_i^{p,h}(M_1M_2) = \bigg(C_i + \frac{C_{i \pm 1}}{N_c}\bigg)N_i^h(M_2) + \frac{C_{i\pm1}}{N_c}\frac{C_F \alpha _s}{4 \pi} \bigg[V_i^h (M_2) + \frac{4 \pi ^2}{N_c} H_i^h (M_1M_2)\bigg] + P_i^{h,p} (M_2),
\eea
where the upper (lower) sign signifies for $i$ odd (even). The superscript $p$ correspond to penguin contributions where it is ommited for $i= 1,2$. The parameters for $h=0$ corrspond to those given in Appendix \ref{A} where P is replaced by V in the final staes PV. And due to the suppressed contributions from positive helicity so only considering the negative helicity amplitudes \citep{Beneke:2006hg}, the LO parameter $N_i$ is given by 
\bea
   N_i^- (M_2) = \begin{cases}
                 0; &   i=\big\{6,8\big\}\\
                 1; &   {\rm otherwise},
                 \end{cases}
\eea
The vertex corrections are given as 
\bea
       V_i^-(M_2)= 
                 \begin{cases}
                 \int _{0}^{1}dx {\rm \Phi _ {b2}}(x) \big[12  {\rm ln} \frac{m_b}{\mu}-18 +g_T(x)\big]; &   {\rm for} \hspace*{.2cm} i=\big \{1-4, 9, 10\big \} \nn\\
                 \int _{0}^{1}dx {\rm \Phi _{a2}}_{M_2}(x) \big[-12  {\rm ln} \frac{m_b}{\mu}+6 -g_T(1-x)\big]; &   {\rm for} \hspace*{.2cm} i=\big\{5,7\big\} \nn\\
                 \int _{0}^{1}dx {\rm \Phi}_{m_2}(x) \big[-6 +h(x)\big]; &   {\rm for} \hspace*{.2cm} i=\big\{6,8\big\}
                 \end{cases}
\eea
The parameter $g_T(x)$ is given as
\bea
g_T(x)=g(x)+\frac{ln x}{1-x},
\eea 
where $g(x)$ is given in the Appendix \ref{A}.
The hard spectator contributions are given by
\bea
H_i^-&&=-\frac{2f_Bf_{V_1}^ \perp}{m_Bm_bF_-^{B\to V_1}(0)}\frac{m_b}{\lambda _B}\int _0^1 dxdy \frac{\phi_1^ \perp (x)\phi _{b2}(y)}{\bar{x}^2y}, \hspace*{1cm} i= \big\{1-4,9,10\big\},\nn\\
H_i^-&&=\frac{2f_Bf_{V_1}^ \perp}{m_Bm_bF_-^{B\to V_1}(0)}\frac{m_b}{\lambda _B}\int _0^1 dxdy \frac{\phi_1^ \perp (x)\phi _{a2}(y)}{\bar{x}^2\bar{y}}, \hspace*{1cm} i= \big\{5,7\big\}, \nn\\
H_i^-&&=\frac{2f_Bf_{V_1}}{m_Bm_bF_-^{B\to V_1}(0)}\frac{m_bm_1}{m_2^2}\frac{m_b}{\lambda _B}\int _0^1 dxdy \frac{\phi_{a1}(x)\phi _{2}^ \perp(y)}{y\bar{x}\bar{y}}, \hspace*{1cm} i= \big\{6,8\big\},
\eea
where we use the divergent integral can be founf finite through the defining parameter as \cite{Beneke:2006hg}
\bea
\int _0^1 dx \frac{\phi _1^ \perp}{\bar{x}^2}=\bigg[\lim_{u\to 1}\frac{\phi _1^ \perp}{\bar{u}}\bigg]X_H^{V_1}+ \int _0 ^1 \frac{dx}{1-x}\bigg[\frac{\phi _1^ \perp(x)}{1-x}-\bigg(\lim _{u \to 1}\frac{\phi _1 ^ \perp(u)}{\bar{u}}\bigg)\bigg],
\eea
where the asymptotic distribution amplitudes are given by
$\phi ^ \perp (x)=6x(1-x), \hspace*{.2cm}\\
 \phi _a (x)=3 (1-x)^2, \hspace*{.2cm} \phi _b (x)= 3x^2$.



\end{appendices}
\end{document}